\documentclass[english]{article}
\usepackage[T1]{fontenc}
\usepackage[latin9]{inputenc}
\usepackage[a4paper]{geometry}
\usepackage{array}
\usepackage{multirow}
\usepackage{amsmath}
\usepackage{amssymb}
\usepackage{setspace}
\PassOptionsToPackage{normalem}{ulem}
\usepackage{ulem}
\makeatletter

\providecommand{\tabularnewline}{\\}

\newcommand{\lyxaddress}[1]{
\par {\raggedright #1
\vspace{1.4em}
\noindent\par}
}

\AtBeginDocument{
  
}

\makeatother

\usepackage{babel}
\begin{document}

\title{Octonion Quantum Chromodynamics}

\author{B. C. Chanyal\textsuperscript{(1)}, P. S. Bisht\textsuperscript{(1)},
Tianjun Li$^{(2)}$ and O. P. S. Negi$^{(1.2)}$%
\thanks{Address from Feb. 21- April 20, 2012:- Institute of Theoretical Physics,
Chinese Academy of Sciences, Zhong Guan Cun East Street 55, P. O.
Box 2735, Beijing 100190, P. R. China.%
}}

\maketitle

\lyxaddress{\begin{center}
\textsuperscript{(1)}\textsuperscript{}Department of Physics\\
 Kumaun University \\
S. S. J. Campus \\
Almora - 263601 (U.K.), India.
\par\end{center}}

\lyxaddress{\begin{center}
\textsuperscript{(2)} Institute of Theoretical Physics\\
Chinese Academy of Sciences\\
 Zhong Guan Cun East Street 55 \\
P. O. Box 2735 \\
Beijing 100190, P. R. China.
\par\end{center}}

\lyxaddress{\begin{center}
Email- bcchanyal@gmail.com\\
ps\_bisht 123@rediffmail.com\\
tli@itp.ac.cn\\
ops\_negi@yahoo.co.in
\par\end{center}}
\begin{abstract}
Starting with the usual definitions of octonions, an attempt has been
made to establish the relations between octonion basis elements and
Gell-Mann $\lambda$ matrices of $SU(3)$ symmetry on comparing the
multiplication tables for Gell-Mann $\lambda$ matrices of $SU(3$)
symmetry and octonion basis elements. Consequently, the quantum chromo
dynamics (QCD) has been reformulated and it is shown that the theory
of strong interactions could be explained better in terms of non-associative
octonion algebra. Further, the octonion automorphism group $SU(3$)
has been suitably handled with split basis of octonion algebra showing
that the $SU(3)_{C}$ gauge theory of colored quarks carries two real
gauge fields which are responsible for the existence of two gauge
potentials respectively associated with electric charge and magnetic
monopole and supports well the idea that the colored quarks are dyons.



\newpage{}
\end{abstract}

\section{Introduction}

\begin{spacing}{1.5}
In spite of the symmetry, conservation laws and gauge fields describe
elementary particle in terms of their field quanta and interactions.
Nevertheless, the role of number system (hyper complex numbers) has
been an important factor in understanding the various theories of
physics from macroscopic to microscopic level. In fact , there has
been a revival in the formulation of natural laws in terms of numbers.
So, according to celebrated Hurwitz theorem there exists \cite{key-1}
four division algebra consisting of $\mathbb{R}$ (real numbers),
$\mathbb{C}$ (complex numbers), $\mathbb{H}$ (quaternions) \cite{key-2,key-3}
and $\mathbb{O}$ (octonions) \cite{key-4,key-5,key-6}. All these
four algebra's are alternative with totally anti symmetric associators.
Real and complex numbers are limited only up to two dimensions, quaternions
are extended to four dimensions (one real and three imaginaries) while
octonions represent eight dimensions ( one real and seven imaginaries).
Real and complex numbers are commutative and associative. Quaternion
are associative but not commutative while its next generalization
to octonions is neither commutative nor associative. Rather, the laws
of alternatively and distributivity are obeyed by octonions. Quaternions
and octonions are extensively used in the various branches of physics
and mathematics. The octonion analysis has also played an important
role in the context of various physical problems \cite{key-7,key-8,key-9,key-10,key-11,key-12,key-13,key-14}
of higher dimensional supersymmetry, super gravity and super strings
etc while the quaternions have an important role to unify \cite{key-15,key-16,key-17}
electromagnetism and weak forces to represent the electroweak $SU(2)\times U(1)$
sector of standard model. Likewise, the octonions are extensively
studied \cite{key-18,key-19} for the description of color quarks
and played an important role for unification programme of fundamental
interactions in terms of successful gauge theories. Furthermore, the
quaternionic formulation of Yang\textendash{}Mill\textquoteright{}s
field equations and octonion reformulation of quantum chromo dynamics
(QCD) has also been developed \cite{key-20} by taking magnetic monopoles
\cite{key-21,key-22,key-23} and dyons (particles carrying electric
and magnetic charges) \cite{key-24,key-25,key-26,key-27} into account.
It is shown that the three quaternion units explain the structure
of Yang-Mill\textquoteright{}s field while the seven octonion units
provide the consistent structure of $SU(3)_{C}$ gauge symmetry of
quantum chromo dynamics. Keeping in view the potential importance
of monopoles and dyons and their possible role in quark confinement,
in this paper, we have made an attempt to construct $SU(3)$ gauge
theory suitably handled with octonions for colored quarks. So, starting
from the usual definitions of octonions, we have established the suitable
connection between octonion basis elements and $SU(3)$ symmetry after
comparing the multiplication tables for Gell-Mann $\lambda$ matrices
of $SU(3$) symmetry and octonion basis elements. Consequently, the
quantum chromo dynamics (QCD) has been reformulated and it is shown
that the theory of strong interactions could be explained better in
terms of non-associative octonion algebra. Further, the octonion automorphism
group $SU(3$) has been reconnected to the split basis of octonion
algebra and it is shown that the $SU(3)_{C}$ gauge theory of colored
quarks describes two real gauge fields identified as the gauge strengths
of two types of chromo charges showing the presence of electric charge
and and magnetic monopoles. So, it is concluded that the present formalism
of colored quarks suitably describes the existence of dyons:particles
carry the simultaneous existence of electric charge and magnetic monopoles.
\end{spacing}

\section{Octonion Definition}

~~~~~~~An octonion $x$ is expressed \cite{key-28,key-29,key-30}
as

\begin{align}
x=(x_{0},\, x_{1},....,\, x_{7})= & x_{0}e_{0}+\sum_{A=1}^{7}x_{A}e_{A}\,\,\,\,\,\,\,\,\,\,\,(A=1,2,.....,7)\label{eq:1}
\end{align}
where $e_{A}(A=1,2,.....,7)$ are imaginary octonion units and $e_{0}$
is the multiplicative unit element. The octet $(e_{0},e_{1},e_{2},e_{3},e_{4},e_{5},e_{6},e_{7})$
is known as the octonion basis and its elements satisfy the following
multiplication rules

\begin{eqnarray}
e_{0}=1,\, & e_{0}e_{A}=e_{A}e_{0}=e_{A},\, & e_{A}e_{B}=-\delta_{AB}e_{0}+f^{ABC}\, e_{C}.\,\,(\forall\, A,B,C=1,2,...,7)\label{eq:2}
\end{eqnarray}
The structure constants $f^{ABC}$ are completely antisymmetric and
take the value $1$. i.e. $f^{ABC}=+1\{\forall(ABC)=(123),\,(471),$
$(257),\,(165),\,(624),\,(543),\,(736)\}$. Here the octonion algebra
$\mathcal{O}$ is described over the algebra of real numbers having
the vector space of dimension $8$. Octonion algebra \cite{key-14}
is non associative and multiplication rules for its basis elements
given by equations (\ref{eq:2}) are then generalized in the following
table as \cite{key-28,key-29,key-30};

\begin{center}
\textbf{\large }%
\begin{tabular}{|c||c||c||c||c||c||c||c|}
\hline 
\textbf{\large $\cdot$} & \textbf{\large $e_{1}$} & \textbf{\large $e_{2}$} & \textbf{\large $e_{3}$} & \textbf{\large $e_{4}$} & \textbf{\large $e_{5}$} & \textbf{\large $e_{6}$} & \textbf{\large $e_{7}$}\tabularnewline
\hline 
\hline 
\textbf{\large $e_{1}$} & \textbf{\large $-1$} & \textbf{\large $e_{3}$} & \textbf{\large $-e_{2}$} & \textbf{\large $e_{7}$} & \textbf{\large $-e_{6}$} & \textbf{\large $e_{5}$} & \textbf{\large $-e_{4}$}\tabularnewline
\hline 
\hline 
\textbf{\large $e_{2}$} & \textbf{\large $-e_{3}$} & \textbf{\large $-1$} & \textbf{\large $e_{1}$} & \textbf{\large $e_{6}$} & \textbf{\large $e_{7}$} & \textbf{\large $-e_{4}$} & \textbf{\large $-e_{5}$}\tabularnewline
\hline 
\hline 
\textbf{\large $e_{3}$} & \textbf{\large $e_{2}$} & \textbf{\large $-e_{1}$} & \textbf{\large $-1$} & \textbf{\large $-e_{5}$} & \textbf{\large $e_{4}$} & \textbf{\large $e_{7}$} & \textbf{\large $-e_{6}$}\tabularnewline
\hline 
\hline 
\textbf{\large $e_{4}$} & \textbf{\large $-e_{7}$} & \textbf{\large $-e_{6}$} & \textbf{\large $e_{5}$} & \textbf{\large $-1$} & \textbf{\large $-e_{3}$} & \textbf{\large $e_{2}$} & \textbf{\large $e_{1}$}\tabularnewline
\hline 
\hline 
\textbf{\large $e_{5}$} & \textbf{\large $e_{6}$} & \textbf{\large $-e_{7}$} & \textbf{\large $-e_{4}$} & \textbf{\large $e_{3}$} & \textbf{\large $-1$} & \textbf{\large $-e_{1}$} & \textbf{\large $e_{2}$}\tabularnewline
\hline 
\hline 
\textbf{\large $e_{6}$} & \textbf{\large $-e_{5}$} & \textbf{\large $e_{4}$} & \textbf{\large $-e_{7}$} & \textbf{\large $-e_{2}$} & \textbf{\large $e_{1}$} & \textbf{\large $-1$} & \textbf{\large $e_{3}$}\tabularnewline
\hline 
\hline 
\textbf{\large $e_{7}$} & \textbf{\large $e_{4}$} & \textbf{\large $e_{5}$} & \textbf{\large $e_{6}$} & \textbf{\large $-e_{1}$} & \textbf{\large $-e_{2}$} & \textbf{\large $-e_{3}$} & \textbf{\large $-1$}\tabularnewline
\hline 
\end{tabular}
\par\end{center}{\large \par}
\begin{description}
\item [{$\qquad\qquad\qquad$~~~~~~~~~~~~~~Table1-}] Octonion
Multiplication table.
\end{description}
Hence, we get the following relations among octonion basis elements
i.e.

\begin{eqnarray}
\left[e_{A},\,\, e_{B}\right] & = & 2f^{ABC}e_{C};\,\,\,\,\,\,\,\left\{ e_{A},\,\, e_{B}\right\} =-\delta_{AB}e_{0};\,\,\,\,\, e_{A}(e_{B}e_{C})\neq(e_{A}e_{B})e_{C};\label{eq:3}
\end{eqnarray}
where brackets $[\,\,]$ and $\{\,\,\}$ are respectively used for
commutation and the anti commutation relations while $\delta_{AB}$
is the usual Kronecker delta-Dirac symbol. Octonion conjugate is thus
defined as,

\begin{align}
\bar{x}= & x_{0}e_{0}-\sum_{A=1}^{7}x_{A}e_{A}\,\,\,\,\,\,\,\,\,\,\,(A=1,2,.....,7).\label{eq:4}
\end{align}
An Octonion can be decomposed in terms of its scalar $(Sc(x))$ and
vector $(Vec(x))$ parts as 

\begin{eqnarray}
Sc(x) & = & \frac{1}{2}(x+\bar{x})=x_{0};\,\,\,\,\,\,\, Vec(x)=\frac{1}{2}(x-\bar{x})=\sum_{A=1}^{7}x_{A}e_{A}.\label{eq:5}
\end{eqnarray}
Conjugates of product of two octonions as well as its own conjugate
are defined as

\begin{eqnarray}
(\overline{xy}) & = & \overline{y}\,\overline{x}\,\,\,;\,\,\,\,\,\,\,\overline{(\bar{x})}\,\,=x;\label{eq:6}
\end{eqnarray}
while the scalar product of two octonions is defined as 

\begin{eqnarray}
\left\langle x\,,\, y\,\right\rangle  & =\sum_{\alpha=0}^{7} & x_{\alpha}y_{\alpha}=\frac{1}{2}(x\,\bar{y}+y\,\bar{x})=\frac{1}{2}(\bar{x}\, y+\bar{y}\, x);\label{eq:7}
\end{eqnarray}
which can be written in terms of octonion units as

\begin{eqnarray}
\left\langle e_{A}\,,\, e_{B}\,\right\rangle  & = & \frac{1}{2}(e_{A}\overline{e_{B}}+e_{B}\overline{e_{A}})=\frac{1}{2}(\overline{e_{A}}e_{B}+\overline{e_{B}}e_{A})=\delta_{AB}.\label{eq:8}
\end{eqnarray}
The norm of the octonion $N(x)$ is defined as

\begin{eqnarray}
N(x)=\overline{x}x & =x\,\bar{x} & =\sum_{\alpha=0}^{7}x_{\alpha}^{2}e_{0};\label{eq:9}
\end{eqnarray}
which is zero if $x=0$, and is always positive otherwise. It also
satisfies the following property of normed algebra

\begin{eqnarray}
N(xy) & =N(x)N(y) & =N(y)N(x).\label{eq:10}
\end{eqnarray}
As such, for a nonzero octonion $x$ , we define its inverse as

\begin{eqnarray}
x^{-1} & = & \frac{\bar{x}}{N(x)}\label{eq:11}
\end{eqnarray}
which shows that

\begin{eqnarray}
x^{-1}x & =xx^{-1} & =1.e_{0};\,\,\,\,(xy)^{-1}=y^{-1}x^{-1}.\label{eq:12}
\end{eqnarray}

\section{SU(3) Generators (Gell-Mann Matrices)}

The Gell-Mann $\lambda$ matrices are used for the representations
of the infinitesimal generators of the special unitary group called
$SU(3)$. This group consists eight linearly independent generators
$G_{A}\forall(A=1,2,3,........8)$ which satisfy the following commutation
relation as

\begin{align}
\left[G_{A}\,,G_{B}\right]= & iF^{ABC}G_{C}\label{eq:13}
\end{align}
where $F^{ABC}$ is the structure constants. It is completely antisymmetric
(i.e. $F^{123}=+1;\; F^{147}=F^{165}=F^{246}=F^{257}=F^{354}=F^{367}=\frac{1}{2}$
and $F^{458}=F^{678}=\frac{\surd3}{2}$ ). Independent generators
$G_{A}(\forall\, A=1,2,3,........8)$ of $SU(3)$ symmetry group are
related with the $3\times3$ Gell-Mann $\lambda$ matrices as

\begin{align}
G_{A}= & \frac{\lambda_{A}}{2}\label{eq:14}
\end{align}
where $\lambda_{A}(\forall\, A=1,2,3,........8)$ are defined as

\begin{align}
\lambda_{1}= & \begin{pmatrix}0 & 1 & 0\\
1 & 0 & 0\\
0 & 0 & 0
\end{pmatrix};\qquad\:\lambda_{2}=\,\begin{pmatrix}0 & -i & 0\\
i & 0 & 0\\
0 & 0 & 0
\end{pmatrix};\nonumber \\
\lambda_{3}= & \begin{pmatrix}1 & 0 & 0\\
0 & -1 & 0\\
0 & 0 & 0
\end{pmatrix};\qquad\lambda_{4}=\,\begin{pmatrix}0 & 0 & 1\\
0 & 0 & 0\\
1 & 0 & 0
\end{pmatrix};\nonumber \\
\lambda_{5}= & \begin{pmatrix}0 & 0 & -i\\
0 & 0 & 0\\
i & 0 & 0
\end{pmatrix};\qquad\lambda_{6}=\,\begin{pmatrix}0 & 0 & 0\\
0 & 0 & 1\\
0 & 1 & 0
\end{pmatrix};\nonumber \\
\lambda_{7}= & \begin{pmatrix}0 & 0 & 0\\
0 & 0 & -i\\
0 & i & 0
\end{pmatrix};\qquad\lambda_{8}=\,\frac{1}{\surd3}\begin{pmatrix}1 & 0 & 0\\
0 & 1 & 0\\
1 & 0 & -2
\end{pmatrix}\label{eq:15}
\end{align}
and satisfy the following properties

\begin{align}
\left(\lambda_{A}\right)^{\dagger}= & \lambda_{A};\nonumber \\
Tr(\lambda_{A})=0\,\,\,\,\,\,\, & Tr(\lambda_{A}\lambda_{B})=2\delta_{AB};\nonumber \\
\left[\lambda_{A},\lambda_{B}\right]= & 2iF^{ABC}\lambda_{C}\,\,(\forall A,B,C=1,2,3.....8).\label{eq:16}
\end{align}
As such, we may summarize the multiplication rules for the generators
(in terms of $\lambda$ matrices) of $SU(3)$ symmetry in the following
table as;

\begin{singlespace}
\begin{center}
\textbf{\large }%
\begin{tabular}{|c||c||c||c||c||c||c||c||c|}
\hline 
\textbf{\large $\cdot$} & \textbf{\large $\lambda_{1}$} & \textbf{\large $\lambda_{2}$} & \textbf{\large $\lambda_{3}$} & \textbf{\large $\lambda_{4}$} & \textbf{\large $\lambda_{5}$} & \textbf{\large $\lambda_{6}$} & \textbf{\large $\lambda_{7}$} & \multirow{1}{*}{\textbf{\large $\lambda_{8}$}}\tabularnewline
\hline 
\hline 
\textbf{\large $\lambda_{1}$} & \textbf{\large $\Lambda_{1}$} & \textbf{\large $i\lambda{}_{3}$} & \textbf{\large $-i\lambda_{2}$} & \textbf{\large $\frac{i}{2}\lambda_{7}$} & \textbf{\large $-\frac{i}{2}\lambda_{6}$} & \textbf{\large $\frac{i}{2}\lambda_{5}$} & \textbf{\large $-\frac{i}{2}\lambda_{4}$} & \textbf{\large $\frac{1}{\surd3}\lambda_{1}$}\tabularnewline
\hline 
\hline 
\textbf{\large $\lambda_{2}$} & \textbf{\large $-i\lambda{}_{3}$} & \textbf{\large $\Lambda_{2}$} & \textbf{\large $i\lambda_{1}$} & \textbf{\large $\frac{i}{2}$$\lambda_{6}$} & \textbf{\large $\frac{i}{2}\lambda_{7}$} & \textbf{\large $-\frac{i}{2}\lambda_{4}$} & \textbf{\large $-\frac{i}{2}\lambda_{5}$} & \textbf{\large $\frac{1}{\surd3}\lambda_{2}$}\tabularnewline
\hline 
\hline 
\textbf{\large $\lambda_{3}$} & \textbf{\large $i\lambda_{2}$} & \textbf{\large $-i\lambda_{1}$} & \textbf{\large $\Lambda_{3}$} & \textbf{\large $-\frac{i}{2}\lambda_{5}$} & \textbf{\large $\frac{i}{2}\lambda_{4}$} & \textbf{\large $\frac{i}{2}\lambda_{7}$} & \textbf{\large $-\frac{i}{2}\lambda_{6}$} & \textbf{\large $\frac{1}{\surd3}\lambda_{3}$}\tabularnewline
\hline 
\hline 
\textbf{\large $\lambda_{4}$} & \textbf{\large $-\frac{i}{2}\lambda_{7}$} & \textbf{\large $-\frac{i}{2}\lambda_{6}$} & \textbf{\large $\frac{i}{2}\lambda_{5}$} & \textbf{\large $\Lambda_{4}$} & \textbf{\large $-\frac{i}{2}\lambda_{3}$} & \textbf{\large $\frac{i}{2}\lambda_{2}$} & \textbf{\large $\frac{i}{2}\lambda_{1}$} & \textbf{\large $-\frac{\surd3}{2}i\lambda_{5}$}\tabularnewline
\hline 
\hline 
\textbf{\large $\lambda_{5}$} & \textbf{\large $\frac{i}{2}\lambda_{6}$} & \textbf{\large $-\frac{i}{2}\lambda_{7}$} & \textbf{\large $-\frac{i}{2}\lambda_{4}$} & \textbf{\large $\frac{i}{2}\lambda_{3}$} & \textbf{\large $\Lambda_{5}$} & \textbf{\large $-\frac{i}{2}\lambda_{1}$} & \textbf{\large $\frac{i}{2}\lambda_{2}$} & \textbf{\large $\frac{\surd3}{2}i\lambda_{4}$}\tabularnewline
\hline 
\hline 
\textbf{\large $\lambda_{6}$} & \textbf{\large $-\frac{i}{2}\lambda{}_{5}$} & \textbf{\large $\frac{i}{2}\lambda_{4}$} & \textbf{\large $-\frac{i}{2}\lambda_{7}$} & \textbf{\large $-\frac{i}{2}\lambda_{2}$} & \textbf{\large $\frac{i}{2}\lambda_{1}$} & \textbf{\large $\Lambda_{6}$} & \textbf{\large $\frac{i}{2}\lambda_{3}$} & \textbf{\large $-\frac{\surd3}{2}i\lambda_{7}$}\tabularnewline
\hline 
\hline 
\textbf{\large $\lambda_{7}$} & \textbf{\large $\frac{i}{2}\lambda_{4}$} & \textbf{\large $\frac{i}{2}\lambda_{5}$} & \textbf{\large $\frac{i}{2}\lambda_{6}$} & \textbf{\large $-\frac{i}{2}\lambda_{1}$} & \textbf{\large $-\frac{i}{2}\lambda_{2}$} & \textbf{\large $-\frac{i}{2}\lambda_{3}$} & \textbf{\large $\Lambda_{7}$} & \textbf{\large $\frac{\surd3}{2}i\lambda_{6}$}\tabularnewline
\hline 
\hline 
\textbf{\large $\lambda_{8}$} & \textbf{\large $-\frac{1}{\surd3}\lambda_{1}$} & \textbf{\large $-\frac{1}{\surd3}\lambda_{2}$} & \textbf{\large $-\frac{1}{\surd3}\lambda_{3}$} & \textbf{\large $\frac{\surd3}{2}i\lambda_{5}$} & \textbf{\large $-\frac{\surd3}{2}i\lambda_{4}$} & \textbf{\large $\frac{\surd3}{2}i\lambda_{7}$} & \textbf{\large $-\frac{\surd3}{2}i\lambda_{6}$} & \textbf{\large $\Lambda_{8}$}\tabularnewline
\hline 
\end{tabular}
\par\end{center}{\large \par}
\end{singlespace}
\begin{description}
\item [{$\qquad\;\qquad\qquad$Table$\,$2-}] Multiplication table for
Gell-Mann $\lambda$ matrices of $SU(3)$ symmetry.
\end{description}
where the $\Lambda_{1},\Lambda_{2},\Lambda_{3},\Lambda_{4},\Lambda_{5},\Lambda_{6},\Lambda_{7},\Lambda_{8}$
are described in terms of $3\times3$ matrices as

\begin{align}
\Lambda_{1}=\Lambda_{2}=\Lambda_{3}=\Lambda_{123}= & \begin{pmatrix}1 & 0 & 0\\
0 & 1 & 0\\
0 & 0 & 0
\end{pmatrix}\Longrightarrow\left(\lambda_{1}\right)^{2}=\left(\lambda_{2}\right)^{2}=\left(\lambda_{3}\right)^{2}=\Lambda_{123};\nonumber \\
\Lambda_{4}=\Lambda_{5}=\Lambda_{45}= & \begin{pmatrix}1 & 0 & 0\\
0 & 0 & 0\\
0 & 0 & 1
\end{pmatrix}\Longrightarrow\left(\lambda_{4}\right)^{2}=\left(\lambda_{5}\right)^{2};\nonumber \\
\Lambda_{6}=\Lambda_{7}=\Lambda_{67}= & \begin{pmatrix}0 & 0 & 0\\
0 & 1 & 0\\
0 & 0 & 1
\end{pmatrix}\Longrightarrow\left(\lambda_{6}\right)^{2}=\left(\lambda_{7}\right)^{2};\nonumber \\
\Lambda_{8}=\frac{4}{3} & \begin{pmatrix}1 & 0 & 0\\
0 & 1 & 0\\
0 & 0 & 1
\end{pmatrix}\Longrightarrow\left(\lambda_{8}\right)^{2}=\frac{4}{3}\widehat{1}.\label{eq:17}
\end{align}
where $\widehat{1}$ is 3$\times$3 unit matrix.

\section{Relation between Octonions and $SU(3)$Generators}

Comparing Table-1 and Table-2, we may observe a resemblance between
the octonions and the Gell-Mann $\lambda$ matrices of $SU(3)$ symmetry
on using simultaneously the relations (\ref{eq:2}) and (\ref{eq:17})
in the following table as \medskip{}

\begin{center}
{\large }%
\begin{tabular}{|c|c|}
\hline 
\textbf{\large Octonions basis} & \textbf{\large SU(3) generators}\tabularnewline
\hline 
\hline 
\textbf{\large $e_{1}$$\longmapsto$} & \textbf{\large $i\lambda_{1}$}\tabularnewline
\hline 
\textbf{\large $e_{2}$$\longmapsto$} & \textbf{\large $i\lambda_{2}$}\tabularnewline
\hline 
\textbf{\large $e_{3}$$\longmapsto$} & \textbf{\large $i\lambda_{3}$}\tabularnewline
\hline 
\textbf{\large $e_{4}$$\longmapsto$} & \textbf{\large $\frac{i}{2}\lambda_{4}$}\tabularnewline
\hline 
\textbf{\large $e_{5}$$\longmapsto$} & \textbf{\large $\frac{i}{2}\lambda_{5}$}\tabularnewline
\hline 
\textbf{\large $e_{6}$$\longmapsto$} & \textbf{\large -$\frac{i}{2}\lambda_{6}$}\tabularnewline
\hline 
\textbf{\large $e_{7}$$\longmapsto$} & \textbf{\large -$\frac{i}{2}\lambda_{7}$}\tabularnewline
\hline 
\textbf{\large $e_{0}$$\longmapsto$} & \textbf{\large $\frac{\surd3}{2}\lambda_{8}$}\tabularnewline
\hline 
\end{tabular}
\par\end{center}{\large \par}
\begin{description}
\item [{$\qquad\;\qquad\qquad$~~~~~~~~~~~~~~Table$\,$3-}] Relation
between Octonion basis and SU(3) generators.
\end{description}
As such, we have the freedom to establish \cite{key-20} a connection
between the octonion basis elements $e_{A}$ and $3\times3$ Gell-Mann
$\lambda$ matrices of $SU(3)$ in the following manner i.e. 

\begin{align}
e_{1}\Rightarrow & i\lambda_{1},\; e_{2}\Rightarrow i\lambda_{2},\; e_{3}\Rightarrow i\lambda_{3}\longmapsto e_{A}\Longleftrightarrow i\lambda_{A};\,\,\,\,\,(\forall\: A=1,2,3);\nonumber \\
e_{4}\Rightarrow & \frac{i}{2}\lambda_{4},\,\,\, e_{5}\Rightarrow\frac{i}{2}\lambda_{5},\Longleftrightarrow e_{A}=\frac{i}{2}\lambda_{A};\,\,\,\,\,(\forall\: A=4,5,);\nonumber \\
e_{6}\Rightarrow & \frac{i}{2}\lambda_{6},\text{\,\,\,}e_{7}\Rightarrow-\frac{i}{2}\lambda_{7},\Longleftrightarrow e_{A}=-\frac{i}{2}\lambda_{A};\,\,\,\,\,(\forall\: A=6,7,);\nonumber \\
e_{0}\Longleftrightarrow & \frac{\surd3}{2}\lambda_{8}.\label{eq:18}
\end{align}
These results are similar to the those derived earlier by Günaydin
Gürsey \cite{key-18} for octonion units and $\lambda$ matrices of
$SU(3)$ symmetry. Here, equation (\ref{eq:18}) satisfies the Cayley
algebra followed by the octonion multiplication rule $e_{A}\centerdot e_{B}=-\delta_{AB}+f_{ABC}e_{C}$.
So, we have the freedom to establish the following relations among
structure constants of octonions and $SU(3)$ symmetry as 

\begin{align}
F^{ABC}\Rightarrow & f^{ABC}\;(\forall\: ABC=123);\nonumber \\
F^{ABC}\Rightarrow\frac{1}{2} & f^{ABC}\;(\forall\: ABC=147,\,246,\,257,\,435,\,516,\,673);\nonumber \\
F^{ABC}\Rightarrow\frac{\surd3}{2} & f^{ABC}\;(\forall\: ABC=458,\,678).\label{eq:19}
\end{align}
\,Hence, we get

\begin{align}
\left[e_{A},e_{B}\right]\Rightarrow & i\left[\lambda_{A},\lambda_{B}\right]\qquad\qquad(\forall\: ABC=123);\nonumber \\
\left[e_{A},e_{B}\right]\Rightarrow & \frac{i}{2}\left[\lambda_{A},\lambda_{B}\right]\qquad\qquad(\forall\: ABC=147,\,246,\,257,\,435,\,516,673);\nonumber \\
\left[e_{A},e_{B}\right]\Rightarrow & \frac{\surd3}{2}i\left[\lambda_{A},\lambda_{B}\right]\qquad\qquad(\forall\: ABC=458,\,678);\label{eq:20}
\end{align}
which are the commutation relations among octonions basis elements
and Gell-Mann $\lambda$ matrices of $SU(3)$ symmetry the so called
Eight fold way. The benefit to write the octonions in terms of Gell-Mann
$\lambda$ matrices of $SU(3)$ symmetry may be described as
\begin{itemize}
\item Non-associativity of octonions does not effect the invariance of the
symmetry group $SU(2)$ spin (or isospin) multiplets for the given
values of structure constants $f^{ABC}$.
\item It is better to describe the $SU(3)$ symmetry in terms of compact
notations of octonions. Accordingly, the theory of strong interactions
could be described better in terms of non associative Cayley algebra.
\item The eighth Gell-Mann $\lambda$ matrix could be designated in terms
of hyper charge which may have the direct link with the scalar octonion
unit $e_{0}$.
\item It may be concluded that the algebra of strong interactions corresponds
to the $SU(3)$ automorphisms of the octonion algebra which is in
support of the results obtained earlier by Günaydin \cite{key-19}. 
\end{itemize}

\section{Octonions and QCD }

The color group $SU(3)$ corresponds to the local symmetry whose gauging
gives rise to Quantum Chromodynamics (QCD).There are two different
types of $SU(3)$ symmetry. The first one is the symmetry that acts
on the different colors of quarks. This symmetry is an exact gauge
symmetry mediated by the gluons. Other $SU(3)$ symmetry is a flavor
symmetry which rotates different flavors of quarks to each other,
or flavor $SU(3$). Flavor SU(3) is an approximate symmetry of the
vacuum of QCD, and is not a fundamental symmetry at all. It is an
accidental consequence of the small mass of the three lightest quarks.
Here, we are interested in exact $SU(3)$ symmetry of colors in terms
of octonion algebra. For this, let us substitute the values of octonion
units $e_{A}$ in terms of $\lambda_{A}$ from equation (\ref{eq:18})
in to equation (\ref{eq:1}) so that we may express an octonion $x$
as

\begin{align}
x= & x_{0}\left(\frac{\surd3}{2}\lambda_{8}\right)+x_{1}\left(i\lambda_{1}\right)+x_{2}\left(i\lambda_{2}\right)+x_{3}\left(i\lambda_{3}\right)+x_{4}\left(\frac{i}{2}\lambda_{4}\right)+x_{5}\left(\frac{i}{2}\lambda_{5}\right)+x_{6}\left(-\frac{i}{2}\lambda_{6}\right)+x_{7}\left(-\frac{i}{2}\lambda_{7}\right)\label{eq:21}
\end{align}
which can further be reduced as

\begin{align}
x= & x_{0}O_{0}+x_{1}\mathcal{O}_{1}+x_{2}\mathcal{iO}_{2}+x_{3}\mathcal{O}_{3}+x_{4}\mathcal{O}_{4}+x_{5}\mathcal{O}_{5}+x_{6}\mathcal{O}_{6}+x_{7}\mathcal{O}_{7}\label{eq:22}
\end{align}
where $\mathcal{O}_{0}\rightarrow\frac{\surd3}{2}\lambda_{8},\,\mathcal{O}_{1}\rightarrow i\lambda_{1},\,\mathcal{O}_{2}\rightarrow i\lambda_{2},\,\mathcal{O}_{3}\rightarrow i\lambda_{3},\,\mathcal{O}_{4}\rightarrow\frac{i}{2}\lambda_{4},\,\mathcal{O}_{5}\rightarrow\frac{i}{2}\lambda_{5},\mathcal{O}_{6}\rightarrow-\frac{i}{2}\lambda_{6},\mathcal{O}_{7}\rightarrow-\frac{i}{2}\lambda_{7}.$
Thus the octonion conjugate be written as as

\begin{align}
\bar{x}= & x_{0}\mathcal{O}_{0}-x_{1}\mathcal{O}_{1}-x_{2}\mathcal{O}_{2}-x_{3}\mathcal{O}{}_{3}-x_{4}\mathcal{O}_{4}-x_{5}\mathcal{O}_{5}-x_{6}\mathcal{O}_{6}-x_{7}\mathcal{O}_{7}.\label{eq:23}
\end{align}
Here the new octonion units $\mathcal{O}_{A}$ associated with the
$SU(3)$symmetry satisfy the octonion algebra

\begin{align}
\mathcal{O}_{A}\centerdot\mathcal{O}_{B}= & -\delta_{AB}+f_{ABC}\,\mathcal{O}_{C}.\label{eq:24}
\end{align}
As such, we may reformulate the theory of strong interactions, the
quantum Chromodynamics (QCD) based on colour $SU(3)_{c}$ whose generators
satisfy the non-associative algebra of octonions. Let us consider
the triplet $(u,d,s)$ (i.e. the up, down, and strange) flavors of
quarks as the three objects of the group namely the $SU(3)$ group
of flavor symmetry. Each quark has been described in terms of three
colors namely the red, blue and green. The dynamics of the quarks
and gluons are controlled by the quantum Chromodynamics Lagrangian.
The gauge invariant QCD Lagrangian is described as 

\begin{align}
\mathcal{L}= & \bar{\psi_{j}}(i\gamma^{\mu}(D_{\mu})_{jk}-m\,\delta_{jk})\psi_{k}-\frac{1}{4}\, G_{\mu\nu}^{a}G_{a}^{\mu\nu}\nonumber \\
= & \bar{\psi_{j}}(i\gamma^{\mu}\partial_{\mu}-m\,)\psi_{j}-g\, G_{\mu}^{a}\bar{\psi_{j}}\gamma^{\mu}T_{jk}^{a}\psi_{k}-\frac{1}{4}\, G_{\mu\nu}^{a}G_{a}^{\mu\nu}\label{eq:25}
\end{align}
where $(j,k=1,2,3)$ are labeled for three quark fields associated
with three colors (namely the red, blue and green) so that we have
\begin{align}
\psi_{j}= & \left(\begin{array}{c}
\psi_{R}\\
\psi_{B}\\
\psi_{G}
\end{array}\right);\qquad\bar{\psi_{j}}=\left(\bar{\psi}_{R},\bar{\psi}_{B},\bar{\psi}_{G}\right)\label{eq:26}
\end{align}
which is a dynamical function of space-time, in the fundamental representation
of the $SU(3)$ gauge group, indexed by $(j,k=1,2,3)$. In equation
(\ref{eq:25}) $G_{\mu}^{a}$ is the octet of gluon fields which is
also a dynamical function of space-time in the adjoint representation
of the $SU(3)$ gauge group, indexed by $a,b,...=1,2,....,8;$ the
$\gamma^{\mu}$ are the Dirac matrices connecting the spinor representation
to the vector representation of the Lorentz group; and $T_{jk}^{a}$
are the generators connecting the fundamental, anti-fundamental and
adjoint representations of the $SU(3)$ gauge group. In our case,
the octonion units connecting to Gell-Mann $\lambda$ matrices provide
one such representation for the generators of $SU(3)$ gauge group.
In equation (\ref{eq:25}), the symbol $G_{\mu\nu}^{a}$ represents
the gauge invariant gluon field strength tensor, analogous to the
electromagnetic field strength tensor $F_{\mu\nu}$ in Electrodynamics.
It is described by

\begin{align}
G_{\mu\nu}^{a}= & \partial_{\mu}G_{\nu}^{a}-\partial_{\nu}G_{\mu}^{a}-g\, f^{abc}\, G_{\mu}^{b}G_{\mu}^{c}\,\,(\forall a,b,c=1,2,3,...,8)\label{eq:27}
\end{align}
where $f^{abc}$ are the structure constants of $SU(3)$ groups as
described above in terms of octonions. In equation (\ref{eq:25}),
the constants $m$ and $g$ control the quark mass and coupling constants
of the theory, subject to renormalization in the full quantum theory.
Here we may introduce a local phase transformation in color space.
Under $SU(3)$ symmetry the spinor $\psi$ transforms as

\begin{align}
\psi\longrightarrow\psi^{'}= & U\psi=\exp\{i\lambda.\alpha(x)\}\psi;\qquad(\lambda=1,2,.....,8)\label{eq:28}
\end{align}
where

\begin{align}
\lambda.\alpha(x)= & \lambda_{1}\alpha_{1}+\lambda_{2}\alpha_{2}+\lambda_{3}\alpha_{3}+\lambda_{4}\alpha_{4}+\lambda_{5}\alpha_{5}+\lambda_{6}\alpha_{6}+\lambda_{7}\alpha_{7}+\lambda_{8}\alpha_{8}\label{eq:29}
\end{align}
which on using Table-3, may directly be written in the in the following
form in terms of octonions i.e.

\begin{align}
\lambda.\alpha(x)= & -ie_{1}\alpha_{1}-ie_{2}\alpha_{2}-ie_{3}\alpha_{3}-2ie_{4}\alpha_{4}-2ie_{5}\alpha_{5}-2ie_{6}\alpha_{6}-2ie_{7}\alpha_{7}+\frac{2}{\surd3}e_{0}\alpha_{8}.\label{eq:30}
\end{align}
As such, the the quantum Chromodynamics (QCD) may be reformulated
in terms of octonions and non-associative algebra in order to explain
its interesting consequences like
\begin{itemize}
\item Quarks confinement
\item Color blindness of nature
\item Asymptotic freedom
\item Calculation for the masses of mesons and baryons etc. 
\end{itemize}

\section{Split octonions}

The split octonions \cite{key-18,key-19,key-20,key-29} are a non
associative extension of split quaternions. They differ from the octonion
in the signature of quadratic form. The split octonion have a signature
$(4,4)$ whereas the octonions have positive signature $(8,0)$. The
Cayley algebra of octonions over the field of complex number is visualized
as the algebra of split octonions with its following basis element,

\begin{align}
u_{0}=\frac{1}{2}\left(e_{0}+ie_{7}\right),\qquad & u_{0}^{\star}=\frac{1}{2}\left(e_{0}-ie_{7}\right),\nonumber \\
u_{j}=\frac{1}{2}\left(e_{j}+ie_{j+3}\right),\qquad & u_{j}^{\star}=\frac{1}{2}\left(e_{j}-ie_{j+3}\right)(\forall j=1,2,3)\label{eq:31}
\end{align}
where $(\star)$ denotes the complex conjugation and $(i=\sqrt{-1})$,
the imaginary unit, commutes with all $e_{A}\,(\forall\, A=1,2,3,....,7)$.
In equation (\ref{eq:31}) $u_{0},\, u_{0}^{\star},\, u_{j},\, u_{j}^{\star}$
are defined as the bi-valued representations of quaternion units $\begin{array}{c}
e_{0}\end{array},\: e_{1},\: e_{2},\: e_{3}$ which satisfy the following \cite{key-31} multiplication rule

\begin{align}
e_{j}e_{k}= & -\delta_{jk}+\epsilon_{jkl}e_{l}\,\,(\forall j,k,l=1,2,3)\label{eq:32}
\end{align}
where $\epsilon_{jkl}$ are the three index Levi-Civita symbols. The
split octonion basis elements $u_{0},\, u_{0}^{\star},\, u_{j},\, u_{j}^{\star}$
satisfy the following multiplication rule

\begin{align}
u_{i}u_{j}=\epsilon_{ijk}u_{k}^{\star};\qquad & u_{i}^{\star}u_{j}^{\star}=-\epsilon_{ijk}u_{k}^{\star}\qquad(\forall\, i,j,k=1,2,3)\nonumber \\
u_{i}u_{j}^{\star}=-\delta_{ij}u_{0};\qquad & u_{i}u_{0}=0;\qquad u_{i}^{\star}u_{0}=u_{i}^{\star}\nonumber \\
u_{i}^{\star}u_{j}=-\delta_{ij}u_{0};\qquad & u_{i}u_{0}^{\star}=u_{0};\qquad u_{i}^{\star}u_{0}^{\star}=0\nonumber \\
u_{0}u_{i}=u_{i};\qquad & u_{0}^{\star}u_{i}=0;\qquad u_{0}u_{i}^{\star}=0\nonumber \\
u_{0}^{\star}u_{i}^{\star}=u_{i};\qquad & u_{0}^{2}=u_{0};\qquad u_{0}^{\star2}=u_{0}^{\star};\qquad u_{0}u_{0}^{\star}=u_{0}^{\star}u_{0}=0.\label{eq:33}
\end{align}
So, we may introduce a convenient realization for the basis elements
$(u_{0},\, u_{0}^{\star},\, u_{j},\, u_{j}^{\star})$ in term of Pauli's
spin matrices as

\begin{align}
u_{0}=\left[\begin{array}{cc}
0 & 0\\
0 & 1
\end{array}\right];\qquad & u_{0}^{\star}=\left[\begin{array}{cc}
1 & 0\\
0 & 0
\end{array}\right];\nonumber \\
u_{j}=\left[\begin{array}{cc}
0 & 0\\
e_{j} & 0
\end{array}\right];\qquad & u_{j}^{\star}=\left[\begin{array}{cc}
0 & -e_{j}\\
0 & 0
\end{array}\right].\qquad(\forall\: j=1,2,3)\label{eq:34}
\end{align}
The Cayley's split octonion algebra may also be expressed via$2\times2$
Zorn's vector matrix realizations as

\begin{align}
A=au_{0}^{\star}+bu_{0}+x_{i}u_{i}^{\star}+y_{i}u_{i}= & \left(\begin{array}{cc}
a & -\overrightarrow{x}\\
\overrightarrow{y} & b
\end{array}\right)\label{eq:35}
\end{align}
where $a$ and $b$ are scalars and $\overrightarrow{x}$ and $\overrightarrow{y}$are
$3-$vectors with the product defined \cite{key-18,key-32} as

\begin{align}
\left(\begin{array}{cc}
a & \overrightarrow{x}\\
\overrightarrow{y} & b
\end{array}\right)\left(\begin{array}{cc}
c & \overrightarrow{u}\\
\overrightarrow{v} & d
\end{array}\right)= & \left(\begin{array}{cc}
ac+\overrightarrow{x}\centerdot\overrightarrow{v} & a\overrightarrow{u}+d\overrightarrow{x}-\overrightarrow{y}\times\overrightarrow{v}\\
c\overrightarrow{y}+b\overrightarrow{v}-\overrightarrow{x}\times\overrightarrow{u} & bd+\overrightarrow{y}\centerdot\overrightarrow{u}
\end{array}\right).\label{eq:36}
\end{align}
Hence, the split octonion conjugate equation may be written via $2\times2$
Zorn's vector matrix realizations as

\begin{align}
\overline{A}=au_{0}+bu_{0}^{\star}-x_{i}u_{i}^{\star}-y_{i}u_{i}= & \left(\begin{array}{cc}
b & \overrightarrow{x}\\
-\overrightarrow{y} & a
\end{array}\right).\label{eq:37}
\end{align}
 So, the norm of $A$ is defined as

\begin{align}
A\overline{A}=\overline{A}A= & (ab+\overrightarrow{x}.\overrightarrow{y})\hat{1}\label{eq:38}
\end{align}
where $\hat{1}$ as the unit matrix of order $2\times2$.

\section{Split-Octonion $SU(3)$ Gauge Theory}

The automorphism group of the Octonion algebra is the 14-dimensional
exceptional $G_{2}$ group that admits a $SU(3)$ subgroup leaving
invariant the idempotents $u_{0}$ and $u_{0}^{\star}$ described
by equation (\ref{eq:31}). This $SU(3)_{C}$ was identified as the
color group acting on the quark and anti-quark triplets \cite{key-19,key-32}.
As such, the automorphism group $SU(3)$ of the quantum mechanical
Hilbert space should be considered as an exact symmetry and can not
be identified as the symmetry of broken unitary spin gauge group.
It is like the $SU(3)_{C}$ color gauge group of quantum chromo-dynamics
(QCD). Therefore, in order to describe the $SU(3)$ gauge theory suitably
handled with split octonions, let us start with the split octonion
equivalent of any four vector $A_{\mu}$ and its conjugate in terms
of the following $2\times2$ Zorn matrix realization as

\begin{align}
Z(A)=\left[\begin{array}{cc}
x_{4} & -\overrightarrow{x}\\
\overrightarrow{y} & y_{4}
\end{array}\right];\qquad & Z(\overline{A})=\left[\begin{array}{cc}
x_{4} & \overrightarrow{x}\\
-\overrightarrow{y} & y_{4}
\end{array}\right].\label{eq:39}
\end{align}
Rather, the Octonion covariant derivative or $\mathcal{O}$- derivative
of an octonion $K$ is defined \cite{key-32,key-33,key--34} as
\begin{eqnarray}
K_{\parallel\mu} & =K_{,\mu}+ & \left[\Im_{\mu},K\right]\label{eq:40}
\end{eqnarray}
where $\Im_{\mu}$ is the Octonion affinity. It is the object that
makes $K_{\parallel\mu}$transform like an octonion under $\mathcal{O}$
transformations i.e.

\begin{eqnarray}
K' & = & U\, K\, U^{-1}\nonumber \\
K'_{\parallel\mu} & = & UK_{\parallel\mu}U^{-1}\nonumber \\
\Im'_{\mu} & = & U\,\Im_{\mu}U^{-1}-\frac{\partial U}{\partial x^{\mu}}\, U^{-1}\label{eq:41}
\end{eqnarray}
where the $U(x)$ are octonions which define local (Octonion) unitary
transformations and are isomorphic to the rotation group $O(3)$.
Thus, equation (\ref{eq:41}) describes $SU(2)$ nature of octonion
$\mathcal{O}$ transformations resulting to the octonion affinity
(gauge potential) $\Im_{\mu}$ of Yang-Mill's type field and is expressed
as.

\begin{eqnarray}
\Im_{\mu} & = & -L_{\mu j}u_{j}^{\star}-K_{\mu j}u_{j}=\left[\begin{array}{cc}
0_{2} & L_{\mu i}e_{j}\\
-K_{\mu j}e_{j} & 0_{2}
\end{array}\right](\forall j=1,2,3)\label{eq:42}
\end{eqnarray}
where the quaternion units $e_{j}=-i\sigma_{j}$ are suitably handled
\cite{key-31} with Pauli spin matrices $\sigma_{j}$. Now, we have
the freedom to extend $SU(2)$ gauge theory to the case of $SU(3)$
Yang Mills gauge theory of colored quarks by replacing the Pauli spin
matrices to Gellmann $\lambda$ matrices. So, from equation (\ref{eq:42}),
it is clear that octonion covariant derivative (\ref{eq:40}) is subjected
by two real ( or one complex) gauge potential transformations. Hence,
$G_{\mu}^{a}$ the octet of gluon fields describing Lagrangian (\ref{eq:25})
is either a complex gauge field or comprises the order pair of two
real gauge fields. So, we may write the covariant derivative $D_{\mu}$
for $SU(3)$ Lagrangian (\ref{eq:25}) as

\begin{align}
D_{\mu}= & \partial_{\mu}+\mathbb{V_{\mu}}\label{eq:43}
\end{align}
where $\mathbb{V_{\mu}}$ is the octonion form of generalized four
potential described as

\begin{alignat}{1}
\mathbb{V_{\mu}} & =e_{0}\left(A_{\mu}^{\alpha}e_{\alpha}\right)+ie_{7}\left(B_{\mu}^{\alpha}g_{\alpha}\right)\qquad(\forall\,\mu=0,1,2,3;\,\alpha=1,2,......,8.)\label{eq:44}
\end{alignat}
The beauty of the equation (\ref{eq:44}) reinforces the $SU(3)$
symmetry of colored quarks with two gauge potentials as the consequence
of automorphism group of split octonions in terms of $2\times2$ Zorn
vector matrix realization. Here, the two gauge potentials $A_{\mu}^{\alpha}$
and $B_{\mu}^{\alpha}$$(\forall\,\mu=0,1,2,3;\,\alpha=1,2,......,8.)$
may be identified as the gauge potentials for two chromo charges supposed
to be responsible for the existence of electric and magnetic chromo-charges.
It may, therefore, be concluded that octonion colored quarks are dyons:
the particles which carry the simultaneous existence of electric and
magnetic charges \cite{key-24,key-25,key-26,key-27}. Substituting
the value of $SU(3)_{C}$octonion gauge potential $\mathbb{V_{\mu}}$
(\ref{eq:44}) in to the equation (\ref{eq:43}), we may write the
covariant derivative $D_{\mu}$ as

\begin{align}
D_{\mu}= & \partial_{\mu}+e_{0}\left(A_{\mu}^{\alpha}e_{\alpha}\right)+ie_{7}\left(B_{\mu}^{\alpha}g_{\alpha}\right)\nonumber \\
= & u_{0}^{*}\left(\partial_{\mu}+A_{\mu}^{\alpha}e_{\alpha}+B_{\mu}^{\alpha}g_{\alpha}\right)+u_{0}\left(\partial_{\mu}+A_{\mu}^{\alpha}e_{\alpha}-B_{\mu}^{\alpha}g_{\alpha}\right)\label{eq:45}
\end{align}
which may is equivalently \cite{key--34} be written as

\begin{align}
D_{\mu}= & \left(\begin{array}{cc}
\partial_{\mu}+\left(e_{\alpha}A_{\mu}^{\alpha}+g_{\alpha}B_{\mu}^{\alpha}\right) & 0\\
0 & \partial_{\mu}+\left(e_{\alpha}A_{\mu}^{\alpha}-g_{\alpha}B_{\mu}^{\alpha}\right)
\end{array}\right).\label{eq:46}
\end{align}
It yields to 

\begin{align}
\left[D_{\mu},D_{\nu}\right]= & \left(\begin{array}{cc}
G_{\mu\nu}^{\alpha}e_{\alpha}+\mathsf{G_{\mu\nu}^{\alpha}}g_{\alpha} & 0\\
0 & G_{\mu\nu}^{\alpha}e_{\alpha}-\mathsf{G_{\mu\nu}^{\alpha}}g_{\alpha}
\end{array}\right)\longmapsto\mathbb{G_{\mu\nu}^{\alpha}}\label{eq:47}
\end{align}
where 

\begin{align}
G_{\mu\nu}^{\alpha}= & \partial_{\mu}A_{\nu}^{\alpha}-\partial_{\nu}A_{\mu}^{\alpha}+e_{\alpha}\left[A_{\mu}^{\alpha},\, A_{\nu}^{\alpha}\right]\longmapsto\, E_{\mu\nu}^{\alpha};\nonumber \\
\mathsf{G_{\mu\nu}^{\alpha}}= & \partial_{\mu}B_{\nu}^{\alpha}-\partial_{\nu}B_{\mu}^{\alpha}+g_{\alpha}\left[B_{\mu}^{\alpha},\, B_{\nu}^{\alpha}\right]\longmapsto\, H_{\mu\nu}^{\alpha};\label{eq:48}
\end{align}
are two $SU(3)$ non-Abelian gauge field strengths in term of electric
$\left(E_{\mu\nu}^{\alpha}\right)$ and magnetic field $\left(H_{\mu\nu}^{\alpha}\right)$
field strengths obtained respectively from electric $A_{\mu}^{\alpha}$
and magnetic $B_{\mu}^{\alpha}$ gauge potentials. Operating the covariant
derivative $D_{\mu}$ (\ref{eq:46}) to the the generalized field
strength $\mathbb{G_{\mu\nu}^{\alpha}}$ of dyons (\ref{eq:47}),
we get

\begin{align}
D_{\mu}\mathbb{G^{\alpha}}_{\mu\nu}= & \left(\begin{array}{cc}
\partial_{\mu}G_{\mu\nu}^{\alpha}e_{\alpha}+\partial_{\mu}\mathsf{G_{\mu\nu}^{\alpha}}g_{\alpha} & 0\\
0 & \partial_{\mu}G_{\mu\nu}^{\alpha}e_{\alpha}-\partial_{\mu}\mathsf{G_{\mu\nu}^{\alpha}}g_{\alpha}
\end{array}\right)=\mathbb{J_{\nu}^{\alpha}}\label{eq:49}
\end{align}
where $\mathbb{J_{\nu}^{\alpha}}$ describes the generalized octonion
gauge current of dyons in term of 2$\times$2 Zorn matrix realization
of split octonion $SU(3)$ gauge theory. It also comprises the electric
and magnetic four currents of dyons as 

\begin{align}
\mathbb{J_{\nu}^{\alpha}}= & \left(\begin{array}{cc}
J_{\nu}^{\alpha}e_{\alpha}+K_{\nu}^{\alpha}g_{\alpha} & 0\\
0 & J_{\nu}^{\alpha}e_{\alpha}-K_{\nu}^{\alpha}g_{\alpha}
\end{array}\right).\label{eq:50}
\end{align}
Here $J_{\nu}^{\alpha}=\partial_{\mu}G_{\mu\nu}^{\alpha}$ and $K_{\nu}^{\alpha}=\partial_{\mu}\mathsf{G_{\mu\nu}^{\alpha}}$
are the four currents respectively associated with the presence of
electric and magnetic charges. So, it is concluded concluded that
split octonion $SU(3)$ gauge theory of colored quarks describes dyons
which are the particles carrying the simultaneous existence of electric
and magnetic monopoles.

\pagebreak{}\textbf{ACKNOWLEDGMENT}: One of us (OPSN) acknowledges
the financial support from Third World Academy of Sciences, Trieste
(Italy) and Chinese Academy of Sciences, Beijing under UNESCO-TWAS
Associateship Scheme. He is also thankful to Prof. Yue-Liang Wu for
his hospitality and research facilities\textbf{ }at ITP and KITPC,
Beijing (China).

\end{document}